# Integrating Genetic Algorithm, Tabu Search Approach for Job Shop Scheduling

R.Thamilselvan[1]
[1]Computer Science and Engineering
Kongu Engineering College
Erode, India
r_thamilselvan@yahoo.co.in

Dr.P.Balasubramanie[2]
[2]Computer Science and Engineering
Kongu Engineering College
Erode, India
pbalu_20032001@yahoo.co.in

*Abstract*— This paper presents a new algorithm based on integrating Genetic Algorithms and Tabu Search methods to solve the Job Shop Scheduling problem. The idea of the proposed algorithm is derived from Genetic Algorithms. Most of the scheduling problems require either exponential time or space to generate an optimal answer. Job Shop scheduling (JSS) is the general scheduling problem and it is a NP-complete problem, but it is difficult to find the optimal solution. This paper applies Genetic Algorithms and Tabu Search for Job Shop scheduling problem and compares the results obtained by each. With the implementation of our approach the JSS problems reaches optimal solution and minimize the makespan.

*Keywords —Genetic Algorithm, Tabu Search, Simulated Annealing, Clustering Algorithm, Job Shop Scheduling.*

## I. INTRODUCTION

An instance of the job-shop scheduling problem [3] consists of a set of **n** jobs and **m** machines. Each job consists of a sequence of **n** activities so there are **nxm** activities in total. Each activity has duration and requires a single machine for its entire duration. The activities within a single job all require a different machine. An activity must be scheduled before every activity following it in its job. Two activities cannot be scheduled at the same time if they both require the same machine. The objective is to find a schedule that minimizes the overall completion time of all the activities.

In this paper, we use the job-shop scheduling. That is, given an instance of the job-shop scheduling problem and completion time **T**, find a schedule whose completion time is less than or equal to **T**.

*Definition:* The Job Shop Scheduling problem is formalized as a set of $n$ jobs (j={$J_1,J_2,\ldots.J_n$}) and a set of $m$ machines (M={$M_1,M_2,\ldots.M_m$}). Each job $J_i$ has $n_i$ subtasks (called *operations* $O_{ij}$), and each operation $J_{ij}$ must be scheduled on a predetermined machine, $P_{ij}$ No machine may process more than one operation at a time, and each operation $J_{ij}\varepsilon\ J_i$ must complete before the next operation in that job ($J_i(j+1)$) begins. Each operation can be processed by only one machine and the performance measure as $C_{max}$.

Table 1 A 3X3 PROBLEM

| Job | Operations Routing (Processing Time) | | | |
|---|---|---|---|---|
| J1 | $M_1(3)$ | $M_2(3)$ | $M_3(3)$ | $M_4(2)$ |
| J2 | $M_3(2)$ | $M_3(3)$ | $M_1(4)$ | $M_2(3)$ |
| J3 | $M_2(3)$ | $M_4(2)$ | $M_2(2)$ | $M_1(4)$ |
| J4 | $M_4(3)$ | $M_1(2)$ | $M_4(2)$ | $M_3(4)$ |

The time required to complete all the jobs is called the *makespan* $C_{max}$. The objective when solving or optimizing this general problem is to determine the schedule which minimizes $C_{max}$. An example of a 4X4 JSSP is given in Table 1. The data includes the routing of each job through each machine and the processing time for each operation (in parentheses). Figure 1 shows a solution for the problem represented by Gantt-Chart.

**Representation models:** A schedule is defined by a complete and feasible ordering of operations to be processed on each machine and in a job shop there are two main ways of graphically representing such an ordering

- Disjunctive graph
- Gantt chart

**Disjunctive graph model:** JSS scheduling problem can be represented by a disjunctive graph. In this disjunctive graph a vertex represents an operation. The conjunctive arcs which are directed lines connect two consecutive operations of the same job. The disjunctive arcs which are pairs of opposite directed lines connect a pair of operations belonging to different jobs to be processed on the same machine. Two additional vertices are drawn to represent the start and the end of a schedule. Let us consider an example of JSS problem with four jobs and three machines, the data are given in Table 1. The disjunctive graph representation for the above example problem is shown in Figure. 2.

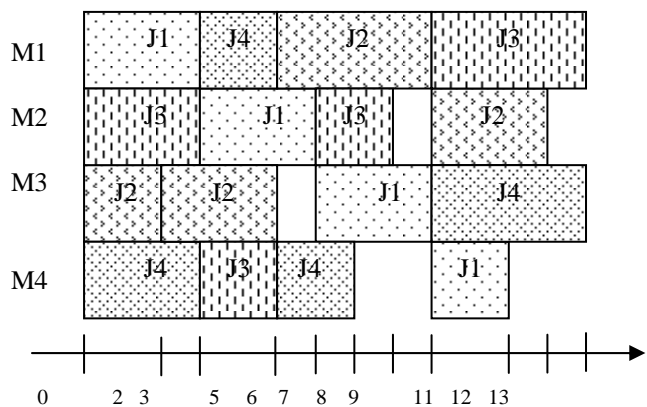

Figure 1: A Gantt-Chart representation of a solution for a 4X4 problem

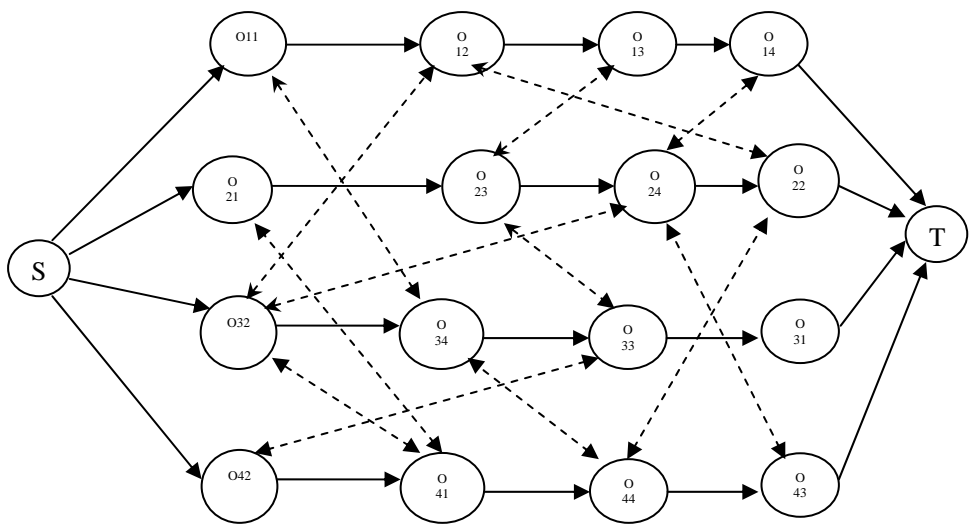

Figure 2: A disjunctive graph of a 4X4 problem

⎯⎯⎯► Conjunctive arc

◄- - - -► Disjunctive arc

Oij: an operation of job i on machine j

Pij: processing time of oij

**Gantt charts:** Gantt chart[5] is the graphical representation of position of jobs and operations on the respective machines. It also represents idle times, starting and completion times of machines. In gantt chart, the various operation blocks are moved to the left as much as possible on each machine and this will help to have a compact schedule which will generally minimize the makespan measure. Figure 1 represents the gantt chart of the feasible solution for the problem in Table 1.

## II. RELATED WORK

*A. Genetic Algorithm (GA)*

GA is an evolutionary technique for large space search. The general procedure of GA search is as follows:

1) Population generation: A population is a set of *chromosomes* and each represents a possible solution, which is a mapping sequence between tasks and machines. The initial population can be generated by other heuristic algorithms, such as Min-min (called seeding the population with a Min-min chromosome).

2) Chromosome evaluation: Each chromosome is associated with a fitness value, which is the makespan of the task-machine mapping this chromosome represents. The goal of GA search is to find the chromosome with optimal fitness value.

3) Crossover and Mutation operation: Crossover operation selects a random pair of chromosomes and chooses a random point in the first chromosome [12]. For the sections of both chromosomes from that point to the end of each chromosome, crossover exchanges machine assignments between corresponding tasks. Mutation randomly selects a chromosome, then randomly selects a task within the chromosome, and randomly reassigns it to a new machine.

4) Finally, the chromosomes from this modified population are evaluated again. This completes one iteration of the GA. The GA stops when a predefined number of evolutions are reached or all chromosomes

Table 2 Genetic Algorithm pseudo code

1. C = set of first operation of each job.
2. t(C) = the minimum completion time among jobs in C.
   - m* = machine where t(C) is achieved.
3. G = set of operations in C that run on m* that and can start before t(C).
4. Select an operation from G to schedule.
5. Delete the chosen operation from C. Include its immediate successor (if one exists) in C.
   - If all operations are scheduled, terminate.
   - Else, return to step 2.

Table 3 Tabu Algorithm pseudo code

```
k := 1.
generate initial solution
WHILE the stopping condition is not met DO
   Identify N(s). (Neighbourhood set)
   Identify T(s,k). (Tabu set)
   Identify A(s,k). (Aspirant set)
   Choose the best s' Î N(s,k) = {N(s) - T(s,k)}+A(s,k).
   Memorize s' if it improves the previous best known solution
   s := s'.
   k := k+1.
END WHILE
```

This genetic algorithm [7] randomly selects chromosomes. Crossover is the process of swapping certain sub-sequences in the selected chromosomes. Mutation is the process of replacing certain sub-sequences with some task-mapping choices new to the current population. Both crossover and mutation are done randomly. After crossover and mutation, a new population is generated. Then this new population is evaluated, and the process starts over again until some stopping criteria are met. The stopping criteria can be, for example, 1) no improvement in recent evaluations; 2) all chromosomes converge to the same mapping; 3) a cost bound is met.

*B. Tabu Search (TS)*

Tabu search is a metaheuristic algorithm that can be used for solving combinatorial optimization problems, such as the Job Shop Scheduling (JSS). Tabu search uses a local or neighbourhood search procedure to iteratively move from a solution $x$ to a solution $x'$ in the neighbourhood of $x$, until some stopping criterion has been satisfied. To explore regions of the search space that would be left unexplored by the local search procedure, tabu search modifies the neighbourhood structure of each solution as the search progresses. The solutions admitted to $N^*(x)$, the new neighbourhood, are determined through the use of memory structures. The search then progresses by iteratively moving from a solution $x$ to a solution $x'$ in $N^*(x)$.

Perhaps the most important type of memory structure used to determine the solutions admitted to $N^*(x)$ is the tabu list. In its simplest form, a tabu list is a short-term memory which contains the solutions that have been visited in the recent past (less than $n$ iterations ago, where $n$ is the number of previous solutions to be stored ($n$ is also called the tabu tenure)). Tabu search excludes solutions in the tabu list from $N^*(x)$. A variation of a tabu list prohibits solutions that have certain attributes (e.g., solutions to the traveling salesman problem (TSP) which include undesirable arcs) or prevent certain moves (e.g. an arc that was added to a TSP tour cannot be removed in the next $n$ moves). Selected attributes in solutions recently visited are labeled "tabu-active.". Solutions that contain tabu-active elements are "tabu". This type of short-term memory is also called "recency-based" memory.

Tabu lists containing attributes can be more effective for some domains, although they raise a new problem. When a single attribute is marked as tabu, this typically results in more than one solution being tabu. Some of these solutions that must now be avoided could be of excellent quality and might not have been visited. To mitigate this problem, "aspiration criteria" are introduced: these override a solution's tabu state, thereby including the otherwise-excluded solution in the allowed set. A commonly used aspiration criterion is to allow solutions which are better than the currently-known best solution.

**Tabu list implementation** Another optimization important to the overall running time of a Tabu search algorithm is the implementation of the Tabu list[1]. While it is convenient to think of this structure as an actual list, in practice, implementing it as such results in a significant amount of computational overhead for all but the smallest lists. Another approach is to store a matrix of all possible operation pairs (i.e. arcs). A time stamp is fixed to an arc when it is introduced into the problem by taking a move, and the time stamping value is incremented after every move. With this representation, a Tabu list query may be performed in constant time (i.e. $currTime-timeStamp_{ij} < length[tabuList]$). Furthermore, the Tabu list may be dynamically resized in constant time.

III. PROPOSED ALGORITHM

*A. Combined Heuristics*

GA can be combined with TS to create combinational heuristics. For example, The Genetic Tabu Search Algorithm (GTA) heuristic is a combination of the GA and TS techniques[1]. In general, GTA follows procedures similar to the GA outlined above. However, for the selection process, GTA uses the Tabu search process.

These Nature's heuristics were only relatively introduced into the scheduling area and more work needs to be done to fit them in a Grid context. There are a lot of interesting questions. First, the meaning of controlling

Table 4 Genetic Tabu Search Algorithm pseudo code

---

1. Central node generates n initial solutions using GA. It runs GA for fixed number of iterations, t.
   a. Choose initial population of fixed size and set j=1
   b. While(j<=t)
      Begin
      i. Apply the operator on the two parent schedules chosen randomly to produce two offspring and replace the parents by the best two the four schedules.
      ii. j=j+1
      End
2. Central node sends *m* best solutions chosen to the *m* remote worker nodes
3. Each worker node runs the TS algorithm by using the initial state received.
4. Upon receiving a converged result from one of the worker nodes, the central node stops execution.

---

measurements in such heuristics need to be refined. For example, each possible pair of tasks, each possible pair of machine assignments, while the other assignments are unchanged.

If the new makespan is an improvement, the new solution is saved, replacing the current solution. Second, there is a trade-off between the search cost and the degree of optimality of solutions found. For example, in a genetic algorithm, historical knowledge can be used to guide the chromosome selection, crossover, or mutation process so that the search process can converge quickly. But this adjustment seems to contradict the philosophy of an evolutionary algorithm: randomization and diversity generate better results, and it may not bring a better solution. Experiments have shown that a good initial solution for TS improves both the quality of the solution as also execution time. We require *m* initial solutions for distributing among *m* nodes; we choose to combine TS with GA.

In GA, an initial population consisting of a set of solution is chosen and then the solutions are evaluated. Relatively more effective solutions are selected to have more off springs which are, in some way, related to the original solutions. If the genetic operator is chosen properly, the final population will have better solutions. GA improves the whole population. TS aim at producing one best solution. For the TS, we require several good initial solutions to ensure the required number of good initial solution. The pseudo-code for the Tabu Search, Genetic Algorithm and Proposed algorithm is given in the Table 2, Table 3 and Table 4 respectively.

## IV. EXPERIMENTS AND RESULTS

Both the above algorithms have been implemented by using Alchemi [12] running on a network of 5 workstations. It is built on a communication channel. The channel is possible to send task messages, result messages, configure messages and partial result messages, amount the various nodes of the network. The full implementation details of the algorithm are given in the following sections.

In the case of Genetic Tabu Search Algorithm (GTA), first the genetic algorithm is run on the central node to get the required *m* initial solutions. These initial solutions are used by the *m* client nodes of the distributed systems as a starting solution for the Tabu Search algorithm. The code that is executed on the central node is the same as the code in Table 4 except that in step 3 the *m* schedules is the best *m* solutions chosen from the population after applying GA. The genetic algorithm starts with an initial population. It then performs the crossover operation and the population is updated. This is repeated a number of times for solving JSS. Since the main focus of this paper is to combine GA withTS, for solving JSS. Each machine can process only one operation at a time. Each job consists of a sequence of operations in a predefined precedence order. The problem is to find a schedule having minimum total time (cost), often called ''makespan'' [9] of the schedule. The algorithm improves on the initial schedule by generating neighborhood schedules and evaluating.

### A. Job Shop Scheduling (JSS)

JSS involved scheduling of various operations relating to a number of jobs on a number of machines. Different techniques exist in the literature for solving JSS. Since the main focus of this chapter is combine Genetic Algorithm and Tabu Search for solving JSS. Each machine can process only one operation at a time. Each job consists of sequence of operations in a predefined precedence order. The problem is to find a schedule having a minimum total time, often called the 'makespan' of the schedule. An initial schedule is obtained for a given set of n jobs to be scheduled on m machines. The GA is applied to the initial schedule. The algorithm improves on the initial schedule by generating neighbourhood schedules and evaluating them. The following assumptions are made for the problem.

- All the jobs are available at time zero
- There are no machine breakdowns
- Operation times of the jobs on the machines are known before hand

A disjunctive graph model with a set of vertices, a set of arcs and a set of edges can be used for representing the problem. The disjunctive graph can be defined as follows

The set of vertices V consists of all the vertices in C and two vertices are demote as S(start) and T(terminate), representing the operation start and end respectively. The set A contains arcs connecting consecutive operations of the same job, as well as arcs from S to the first operation of each job and from the last operating of each job to T. The edges in the set E connect operations to be processed by the same machine.

*B. Performance of a Proposed Algorithm*

In order to prove the efficiency of Genetic Algorithm with a Tabu Search, we have done a series of experiments for Job Shop Scheduling. The performance of the GA, TS and proposed algorithm are compared for different sizes of the JSS problem (from 5 jobs and 5 machines to 10 jobs on 10 machines). The performance of the algorithms has been compared on the basis of two factors, namely the, the execution time of the algorithm and the cost of the solution, It can be observed from Table 5 and Figure 3 that GTA performed very well in terms of both execution time and the quality of the solution as compared to the other algorithms.

## V. CONCLUSION AND FUTURE WORK

The Genetic Algorithm was implemented for the Job Shop Scheduling as a first application and it was introduced by Nakano and Yamada using a bit string representation and conventional genetic operators. This approach is simple and easy. But it is not very powerful. The same idea is also proposed by Dorndorf and Pesch. The ideas of both groups and other active schedule-based GAs are suitable for middle-size problems; however, it seems necessary to combine each with other heuristics such as the shifting bottleneck or local search to solve larger-size problems. To solve larger-size problems the Tabu Search and Genetic Algorithms are combined together.

In this paper we have presented a Genetic Tabu Search Algorithm for the Job Shop Scheduling (JSS). The GTA algorithm is compared with the Genetic Algorithm and Tabu Search algorithm. The Result shows that the GTS is better than the existing problem GA, TS. The algorithm showed a very good result for the number of nodes increased.

The algorithm is tested in network of homogeneous work stations. In Future the GTS algorithm is experimented with the dynamic switching from GA to TS and also the same algorithm will be implemented in network of heterogeneous work stations.

Table 5 Comparison of GTA and TS performance for the TSP

| Problem size (jobsXm/cs) | GA | | TS | | GTA | |
|---|---|---|---|---|---|---|
| | Time (s) | Cost | Time (s) | Cost | Time (s) | Cost |
| 5X5 | 255 | 902 | 455 | 850 | 210 | 850 |
| 5X10 | 155 | 1550 | 300 | 1476 | 100 | 1080 |
| 10X5 | 350 | 902 | 550 | 855 | 280 | 910 |
| 5X10 | 140 | 1400 | 280 | 1280 | 120 | 980 |

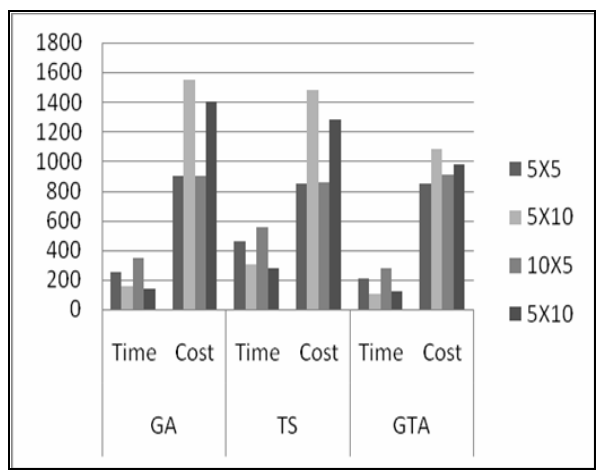

Figure 3 Comparison of GTA and TS performance for the TSP

REFERENCES

[1]. R.Thamilselvan and Dr.P.Balasubramanie, A Genetic Algorithm with a Tabu Search (GTA) for Travelling Salesman Problem. International Journal of Recent Trends in Engineering, Issue. 1, Vol. 1, pp. 607-610, June 2009.
[2]. K. Aggarwal and R. D. Kent, *An Adaptive Generalized Scheduler for Grid Applications*, in Proc. of the 19th Annual International Symposium on High Performance Computing Systems and Applications (HPCS'05), pp.15-18, Guelph, Ontario Canada, May 2005.
[3]. M, Arora, S.K. Das, R. Biswas, *A Decentralized Scheduling and Load Balancing Algorithm for Heterogeneous Grid Environments*, in Proc. of International Conference on Parallel Processing Workshops (ICPPW'02), pp.:499 – 505, Vancouver, British Columbia Canada, August 2002.
[4]. R. Bajaj and D. P. Agrawal, *Improving Scheduling of Tasks in A Heterogeneous Environment*, in IEEE Transactions on Parallel and Distributed Systems, Vol.15, no. 2, pp.107 – 118, February 2004.
[5]. J Blythe, S Jain, E Deelman, Y Gil, K Vahi and A Mandal,K Kennedy, *Task Scheduling Strategies for Workflow-based Applications in Grids*, in Proc. Of International Symposium on Cluster Computing and Grid (CCGrid'05), pp.759-767, Cardiff, UK, May 2005.
[6]. Greening, Daniel R., "Parallel Simulated Annealing Techniques", Physica D, Vol.42, pp. 293-306,1990
[7]. Van Laarhoven, P.J.M., E.H.L., AArts, and Jan Karel Lenstra, "Job Shop Scheduling by Simulated Annealing", Operation Research, Vol. 40, pp. 113-125,1992.
[8]. Syswerda, Gilbert, "Schedule Optimization Using Genetic Algorithms", L. Davis(ed.), Handbook of Genetic Algorithmsm pp.332-349,1991.
[9]. Nowicki, E. and C.Smutnicki, "A Fast Tabu Search Algorithm for Job Shop Problem", Report 8/93, Institute of Engineering Cybernetics, Technical University of Wroclaw, 1993.
[10]. Whitley, Darrel, Timothy, Starkweather and Daniel, Shaner, "Schedule Optimization Using Genetic Algorithms", Lawrence Davis, (ed.), pp.351-357
[11]. Abraham, R. Buyya and B. Nath, Nature's Heuristics for Scheduling Jobs on Computational Grids, in Proc. of 8th IEEE International Conference on Advanced Computing and Communications (ADCOM 2000), pp. 45-52, Cochin, India, December 2000.
[12]. Andresen, S. Kota, M. Tera, and T. Bower. An ip-level network monitor and scheduing system for clusters. In *Proceeding of the 2002 International Conference on Parallel and Distributed Processing Techniques and Applications (PDPTA'02)*, LasVegas.
[13]. Adams, J., Balas, E., Zawack, D., 1988. The shifting bottleneck procedure for job shop scheduling. Management Science 34, pp. 391–401.
[14]. Anderson, E.J., Glass, C.A., Potts, C.N., 1997. Machine scheduling. In: Aarts, E.H.L., Lenstra, J.K. (Eds.), Local Search Algorithms in Combinatorial Optimization. Wiley, pp. 361–414.
[15]. John, D.J., 2002. Co-evolution with the Bierwirth–Mattfeld hybrid scheduler. In: Proceedings of the GECCO 2002 Conference, New York.
[16]. G. Vivo-Truyols, J.R. Torres-Lapasio, M.C. Garcia-Alvarez- Coque, Chemom. Intell. Lab. Syst. 59 (2001) 89–106.
[17]. F. Glover, M. Laguna, Tabu Search, Kluwer Academic Publishers, Dordrecht, 1998.
[18]. Geyik.F and I.H.Cedimoglu, 2004. The Strategies and Parameters of Tabu Search for Job Shop Scheduling, J.Intelligent Manufacturing, pp 439-448

[19]. Glover.F., 1989. Tabu Search-Part I, Operatons Research Society of America. J.Comput, pp 4-32.
[20]. Most popular Grid computing web site www.gridbus.org.
[21]. Most related articles web page www.buyya.com.

AUTHORS PROFILE

R.Thamilselvan is an Assistant Professor in the department of computer Science and Engineering, Kongu Engineering College, Perundurai, Tamilnadu India. He has completed his M.E Computer Science and Engineering in 2005 under Anna University Chennai. He has completed 8 years of teaching service. He has published 3 papers in national conference and 1 paper in International Journal. He was the recipient of the Best Faculty award during the year 2007-2008. His area of interest includes Grid Computing, Parallel Processing, and Distributed Computing. He has organized 2 national level seminar sponsored by AICTE, New Delhi.

Dr.P.Balasubramanie is a Professor in the department of computer Science and Engineering, Kongu Engineering College, Perundurai, Tamilnadu India. He was awarded Junior research Fellowship by Council of Scientific and Industrial Research(CSIR) in 1990 and he has completed his Ph.D degree in 1990 under Anna University in 1996. He has also qualified for the state Level Eligibility test for Lectureship in 1990. He has completed 13 years of teaching service. He has published more than 60 articles in International/National Journals. He has authored six books with the reputed publishers. He was the recipient of the Best Faculty award for consecutively two years. He is also the recipient of the CTS Best Faculty Award-2008. He has guided 3 Ph.D scholars and 23 research scholars are working under his guidance. His area of interest includes Image processing, datamining, networking and so on. He is a member of Board of studies, Anna University Coimbatore. He has organized several seminar/workshops.